*We describe cyberattack surfaces and potential solutions for securing connected and automated vehicles and transportation infrastructure.*

# Security of Connected and Automated Vehicles

Mashrur Chowdhury, Mhafuzul Islam, and Zadid Khan

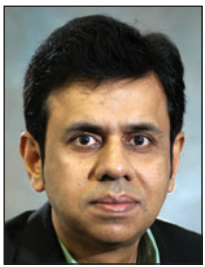 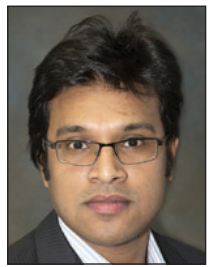 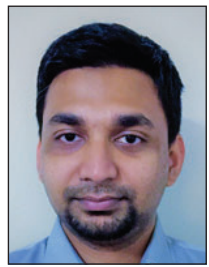

Mashrur Chowdhury    Mhafuzul Islam    Zadid Khan

The transportation system is rapidly evolving with new connected and automated vehicle (CAV) technologies that integrate CAVs with other vehicles and roadside infrastructure in a cyberphysical system (CPS). Through connectivity, CAVs affect their environments and vice versa, increasing the size of the cyberattack surface and the risk of exploitation of security vulnerabilities by malicious actors. Thus, greater understanding of potential CAV-CPS cyberattacks and of ways to prevent them is a high priority.

In this article we describe CAV-CPS cyberattack surfaces and security vulnerabilities, and outline potential cyberattack detection and mitigation strategies. We examine emerging technologies—artificial intelligence,

Mashrur Chowdhury is the Eugene Douglas Mays Professor of Transportation, director of the Center for Connected Multimodal Mobility (C²M²), and codirector of the Complex Systems, Analytics and Visualization Institute, and Mhafuzul Islam and Zadid Khan are PhD students, all in the Glenn Department of Civil Engineering at Clemson University.

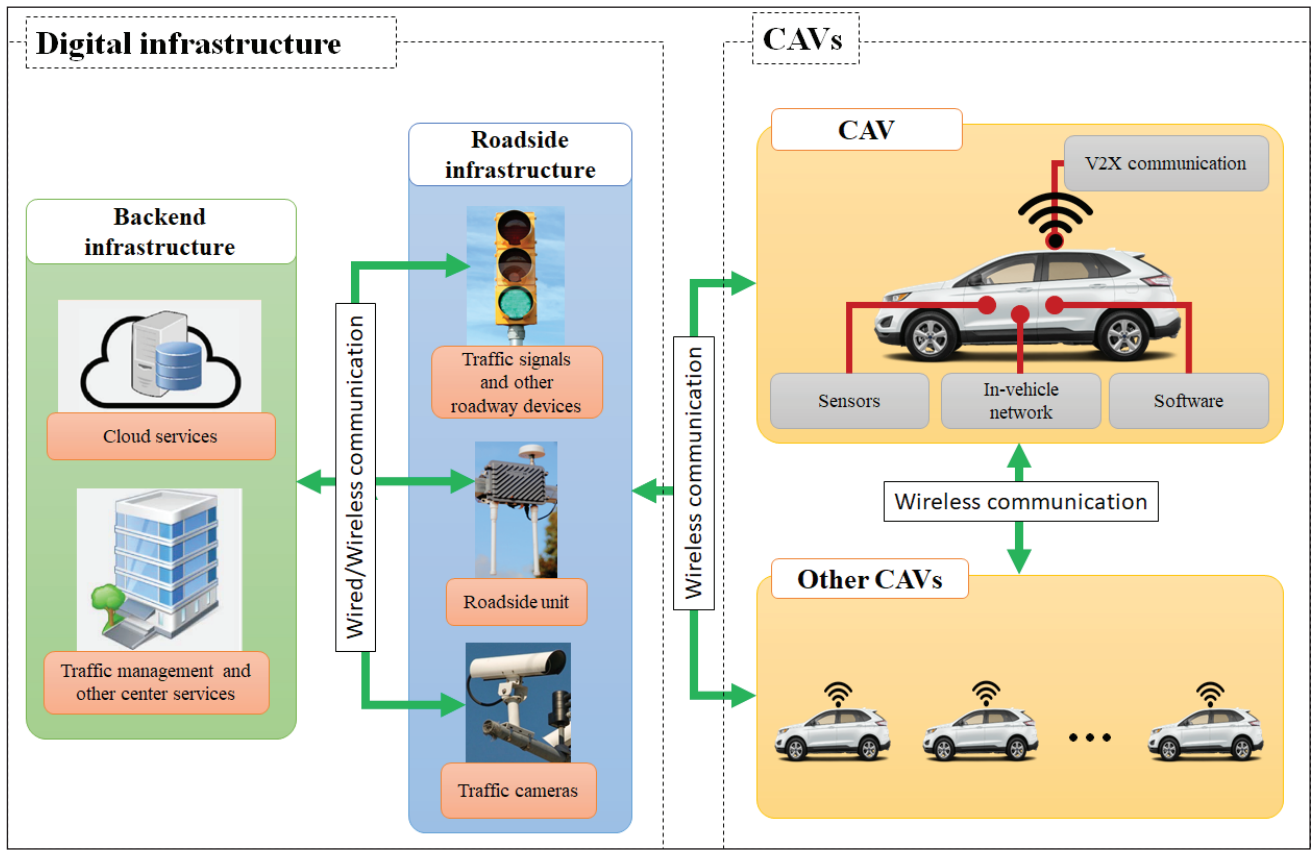

FIGURE 1 Overview of a cyberphysical system for connected and automated vehicles (CAVs). V2X = vehicle-to-everything.

software-defined networks, network function virtualization, edge computing, information-centric and virtual dispersive networking, fifth generation (5G) cellular networks, blockchain technology, and quantum and postquantum cryptography—as potential solutions aiding in securing CAVs and transportation infrastructure against existing and future cyberattacks.

**Introduction**

CAVs are the subject of research and development by both academia and industry because of their potential to improve traffic safety and operations.

The limitations of human perception prevent motorists from discerning what is beyond the range of human sight, such as roadway incidents and work zones around a corner or in the far distance. Even autonomous vehicles' sensors fail to discern many such impediments. Vehicle-to-everything (V2X) communication provides a 360-degree view that transcends the limited capabilities of both human-driven and automated vehicles.

CAVs are planned to be part of a broader connected city initiative, communicating with other CAVs and with smart infrastructure and services (figure 1). CAV-CPS will use connectivity to improve roadway operational efficiency using real-time roadway traffic information (e.g., about traffic signal phasing and timing, traffic incidents and queues) while improving safety in numerous ways, such as redundancy in case an automated vehicle's onboard sensors fail.

However, the highly interconnected CAV-CPS will introduce challenging security issues and vulnerabilities (Deka et al. 2018), with far-reaching consequences for a poorly secured system. For example, attackers could gain access to CAV control systems and cause catastrophic multivehicle crashes. Thus, it is critical to develop security solutions to protect CAVs, their occupants, other road users, and the associated infrastructure.

The Society of Automotive Engineers has created a cybersecurity guidebook of recommended practice, SAE J3061, that establishes a common terminology for security threats, vulnerabilities, and risks across the vehicular CPS (SAE 2016). SAE J3061 recommended practices are based on ISO 26262, an established standard for automotive functional safety (ISO 2018). The rapid evolution of CAV technologies requires adapting these standards and developing new ones to address CAV-CPS security challenges.

TABLE 1 Examples of potential attacks and countermeasures on CAV sensors

| Sensor type | Attack type | Attack description | Potential countermeasures |
|---|---|---|---|
| GPS | Jamming | Blocking or interfering with the satellite signals received by the GPS | • Multiple GPS receivers and GPS activity monitoring (Parkinson et al. 2017)<br>• Antijamming devices as GPS frequency filters (Chien 2015) |
| | Spoofing | Modifying the position, velocity, and time values of the GPS receiver | • Navigation message authentication and cross-correlation of multiple GPS receivers (Psiaki et al. 2013)<br>• Integrated navigation systems combining GPS and inertial navigation system (Jwo et al. 2013)<br>• GPS data validation using V2X communication (Anouar et al. 2017) |
| Camera | Illusion and blinding | Adversarial physical objects compromising the vision-based systems | • Redundant sensors to verify camera data (Petit et al. 2015)<br>• Machine learning models trained with adversarial data (Islam et al. 2019) |
| LiDAR | Jamming | Turning off or degrading its performance | • Protective glasses around a LiDAR that act as light filters (Tuchinda et al. 2006) |
| | Spoofing | Tricking it into detecting false objects | • Misbehavior detection system incorporating other sensor data, such as from the camera (Petit et al. 2015) |

**CAV Security Vulnerabilities and Solutions**

We divide CAV-CPS cyberattack surfaces into three main groups—

- in-vehicle systems, which include sensors, software, and in-vehicle network;
- V2X communication networks; and
- supporting digital infrastructure—

and outline countermeasures to cyberattacks in each category.

*In-Vehicle Systems*

Given the fatal consequences that may result if a CAV's in-vehicle systems are compromised, ensuring their security is of paramount importance to the automotive industry.

*Sensors*

In addition to V2X systems, a CAV is generally equipped with light detection and ranging (LiDAR), camera, radio detection and ranging (RADAR), global positioning system (GPS) and inertial measurement unit sensors. They collaborate to improve vehicle safety and operational efficiency and are responsible for localization, obstacle avoidance, and trajectory and path planning (Schwarting et al. 2018). Table 1 presents examples of potential cyberattacks on GPS, camera, and LiDAR, as automated vehicle safety and operation largely depend on these sensors.

*Software*

Modern vehicles are equipped with multiple electronic control units (ECUs) to manage vehicle functionality through signal acquisition, processing, and control. ECUs may have vulnerabilities that can be exploited by an attacker, and CAVs, with their additional sensors and functionalities, have more ECUs than non-CAVs do (Wyglinski et al. 2013). One such additional ECU is the navigation control module; if compromised, it may misdirect the vehicle toward an unintended destination or even off the road (Chattopadhyay and Lam 2018).

Because ECUs are interdependent, if one is compromised, others are affected. For example, an attack on the engine control module may result in the transmission of false data about wheel speed to the electronic brake control module, which may inappropriately activate the brakes (Parkinson et al. 2017).

One of the biggest security concerns with CAV software is the over-the-air update, which may allow malware injections and thus enable an attacker to gain remote access and control of the vehicle (Nie et al. 2017). Another security threat is software bugs in the source code of the ECU software/firmware.

A simple way to ensure software/firmware security is to use strong cryptographic solutions, which include message authentication, to prevent malware injections (Pike et al. 2017). Signature- and behavior-based models can secure the communication between ECUs (Parkinson et al. 2017). However, using secured over-the-air updates, these mitigation strategies can be kept up to date to maintain and strengthen the security of the CAV software/firmware (Mayilsamy et al. 2018).

*In-Vehicle Network*

ECUs in a CAV communicate with each other using communication message protocols, such as FlexRay, or controller area network (CAN). CAN is a universal real-time messaging protocol widely used by the automotive industry given its lower implementation cost (Liu et al. 2017).

Although the CAN bus has some basic security features, such as firewalls, it is vulnerable to cyberattacks. The absence of authentication and encryption allows unauthorized devices to join the CAN through an onboard device port. As a broadcast-based network, CAN messages have neither source nor destination address, which means that every in-vehicle device can listen to any unencrypted CAN message. The CAN bus is also vulnerable to denial of service (DoS) attacks. For example, an attacker generating high-priority CAN messages can prevent ECUs from communicating with each other because of the prioritization rule of the CAN protocol (Koscher et al. 2017).

Another alternative in-vehicle communication system is the automotive Ethernet, which provides a higher data transmission rate, better reliability and adaptability (Huo et al. 2015), and more security than the CAN (because the Ethernet protocol contains source and destination addresses) (Jadhav and Kshirsagar 2018). In addition, the highly secure hardware security module, trusted platform module (Corbett et al. 2018), and hardware authentication (Intel 2019) are viable solutions for in-vehicle network security.

*Vehicle-to-Everything Network*

In a V2X network, to communicate with external agents such as pedestrians, other vehicles, roadside transportation infrastructure, and servers (cloud-based or in-house), a CAV uses vehicle-to-vehicle (V2V) and vehicle-to-infrastructure (V2I) communication. Cyberattacks on the V2X network can compromise its availability, integrity, confidentiality, and authenticity (Alnasser et al. 2019). Following are the most likely attacks on a V2X network:

- Black- and greyhole attacks: The compromised vehicle stops forwarding all packets (blackhole) or some packets (greyhole) to other CAVs, so that other CAVs cannot receive safety-critical information, such as forward collision warnings.
- DoS and distributed DoS (DDoS): Attackers disrupt the V2X network by data flooding, causing a delay in the transmission of safety-critical information and making the network services unavailable.
- Jamming: An attacker broadcasts signals to corrupt data or disable communication channels.
- False message injection: The compromised vehicle creates fake messages or alters received messages and broadcasts them.
- Eavesdropping: The compromised vehicle uses false identities to capture data packet information, thus acquiring sensitive and confidential data.
- Certificate replication: An attacker conceals itself from certification authorities by replicating the certificates of legitimate vehicles.
- Sybil attack: An attacker creates multiple identities to gain the trust of legitimate CAVs.
- Impersonation: An attacker establishes itself as trustworthy in the network by impersonating a trusted entity to gain access to sensitive information.

Solutions for V2X network security can be based on cryptography, behavior, or identity.

- Cryptography-based solutions include encryption, secure key management and authentication. For example, a security credential management system can prevent impersonation, false identity, and eavesdropping attacks (Ravi and Kulkarni 2013; Whyte et al. 2018).
- Among behavior-based models, one of the most popular is the weighted-sum method, in which a level of trust of a connected vehicle is based on the weighted sum of relevant criteria, such as the transmission range, vehicle speed, and vehicle direction (Sugumar et al. 2018). For example, a blackhole attack can be detected by a reputation-based global trust model, where a CAV's past behavior is considered in determining its current trust value (Li et al. 2013). Behavior-based cooperative awareness between vehicles can prevent greyhole attacks (Ali

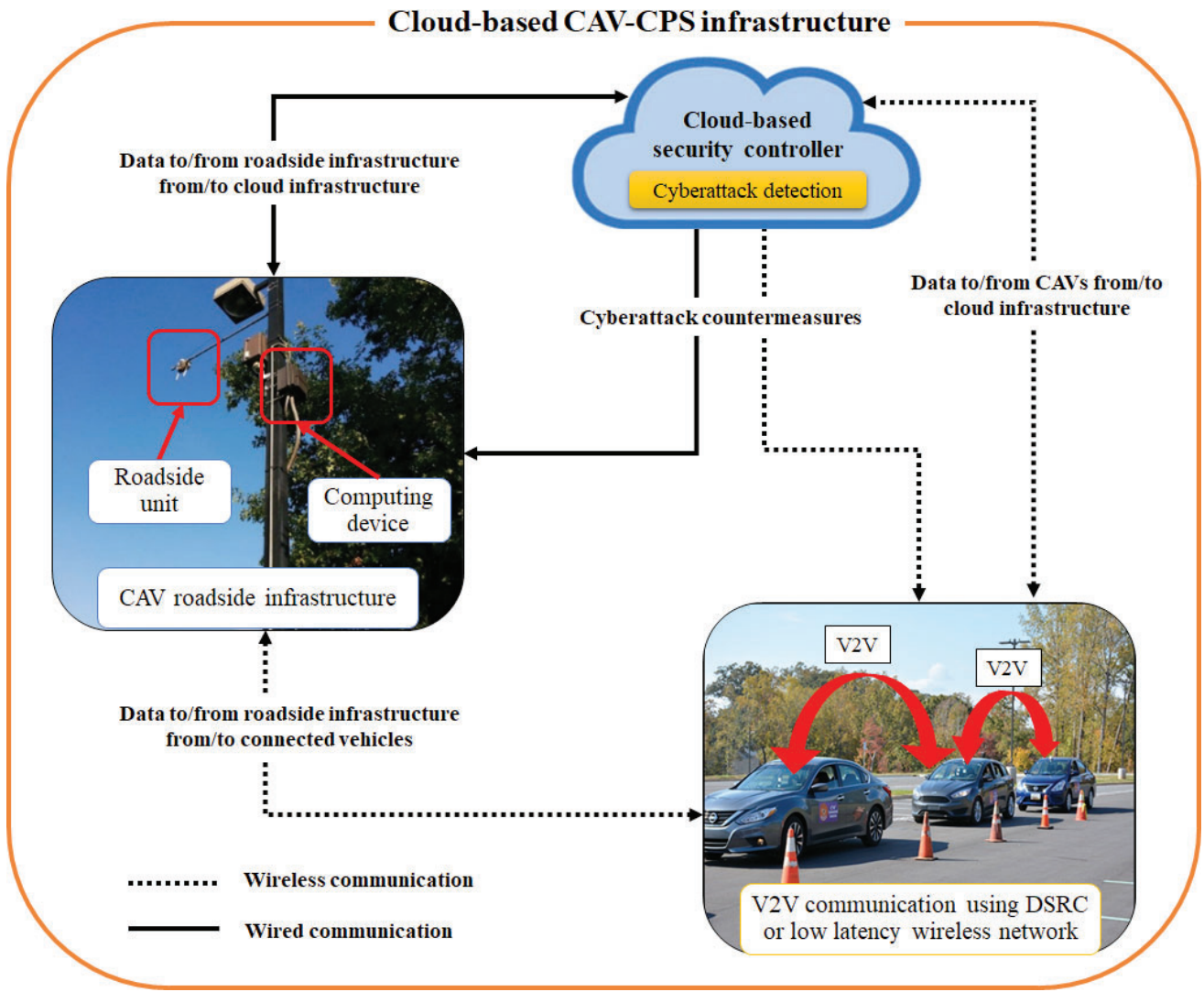

FIGURE 2  Cloud-based security for a connected and automated vehicle (CAV)–cyberphysical system (CPS) at the Center for Connected Multimodal Mobility ($C^2M^2$) testbed, Clemson, SC. DSRC = dedicated short-range communications; V2V = vehicle-to-vehicle.

Alheeti et al. 2016). DoS and DDoS attacks can be detected and mitigated using policy- and rule-based techniques (Islam et al. 2018).

- Identity-based security solutions can be used to detect eavesdropping and secure vehicle privacy (Kang et al. 2016).

*Digital Infrastructure*

In addition to V2X security, the security of the CAV-CPS digital infrastructure is of major concern. Digital infrastructure includes roadside units, traffic signals and cameras, traffic management centers, and cloud infrastructure (figure 1). Digital infrastructure enables cloud-based security solutions for different components of the V2X network and CAVs (figure 2), such as

- connected vehicles communicating with each other using dedicated short-range communications or any other low-latency wireless network;
- roadside infrastructure communicating with connected vehicles via a low-latency wireless network and other roadside infrastructure using wireless/wired networks;
- computing devices running connected vehicle applications;
- cloud servers hosting a security controller for detecting and mitigating cyberattacks to the V2X network.

We used a cloud-based attack detection and mitigation system to detect cyberattacks on the V2X network and roadside infrastructures in a testbed at the Center for Connected Multimodal Mobility ($C^2M^2$), a US Department of Transportation (USDOT) Tier 1 University Transportation Center headquartered in Clemson, South Carolina (Islam et al. 2018). Although the cloud would add some vulnerabilities to the system, cloud security is well studied, and protective techniques are available (Basu et al. 2018). However, an attacker can take advantage of the vulnerabilities of the protocols associated with message transfers between the CAVs/roadside infrastructure and cloud servers to cut off the communication link between them (Dinculeană and Cheng 2019). Distributed security solutions using the software-defined network can solve this issue by giving autonomy to the CAVs or roadside infrastructure in terms of operating their corresponding security solutions independently (Darabseh et al. 2015).

**Future Landscape for CAV Cybersecurity**

The rapid evolution of information technology has led to the development and use of emerging technologies to enhance CAV security. Although all emerging technologies are not fully tested and have their own challenges, their potential for securing CAVs is beyond dispute. When creating CAV security solutions using these emerging technologies, a security-by-design process that integrates cyberattack detection and countermeasures needs to be adopted at the outset (Chattopadhyay and Lam 2018).

*Artificial Intelligence*

Advances in computer hardware technologies and distributed computing facilities (e.g., cloud computing) have led to numerous innovations in AI-based cybersecurity that can be used to protect CAVs.

Deep neural networks (DNNs) are used in the development of intrusion detection systems (IDS) for connected vehicles (Aloqaily et al. 2019). Recurrent neural networks, such as a long short-term memory, are perhaps the most relevant and widely used deep learning model for IDS (Levi et al. 2018). Another use of AI for security is context-aware user-behavior analytics, in which CAV behavior data are collected and used to detect security threats (Wasicek et al. 2017).

However, AI itself can be subject to cyberattacks. For example, by placing a small amount of graffiti or a sticker on a road traffic sign, an attacker can cause the DNN-based traffic sign recognition system of an automated vehicle to misclassify the sign, which can lead to traffic crashes (Eykholt et al. 2018). The attacker specifically targets the DNN to compromise the performance of the automated vehicle even as the changes to the sign are imperceptible to humans (Papernot et al. 2017).

When designing a CAV security solution using AI, the AI system itself must be secured by improving its robustness.

> *When designing a CAV security solution using AI, the AI system itself must be secured by improving its robustness.*

*Software-Defined Network*

Traditional network management is nonprogrammable, complex, and error-prone (Modieginyane et al. 2018), but a programmable software-defined network (SDN) can be organized for more secure network management and enhanced CAV security (Nobre et al. 2019).

SDN involves a centralized controller for the data flow between vehicles and the roadside infrastructure (Jaballah et al. 2019), making it an important tool for CAV security. During a cyberattack, the roadside infrastructure routes the data to the SDN controller, and a detection and mitigation application in cyberspace detects the attack and selects an appropriate mitigation strategy (figure 3). An SDN controller can push the mitigation strategy to SDN-enabled switches (e.g., OpenFlow switches) in the physical space to deploy the updated mitigation strategy. Furthermore, SDN combined with AI can be used to detect and mitigate cyberattacks on the in-vehicle system (Khan et al. 2019).

*Network Function Virtualization*

Unlike legacy systems, in which network functions, such as firewalls and IDS, are deployed in proprietary hardware, network function virtualization (NFV) can provide a cost-effective approach to CAV security (Alwakeel et al. 2018). Virtualized network functions can run on top of commodity/off-the-shelf hardware, such as industry-standard servers, storage, and switches. Virtualization of network security functions can be used to design security solutions that provide the same

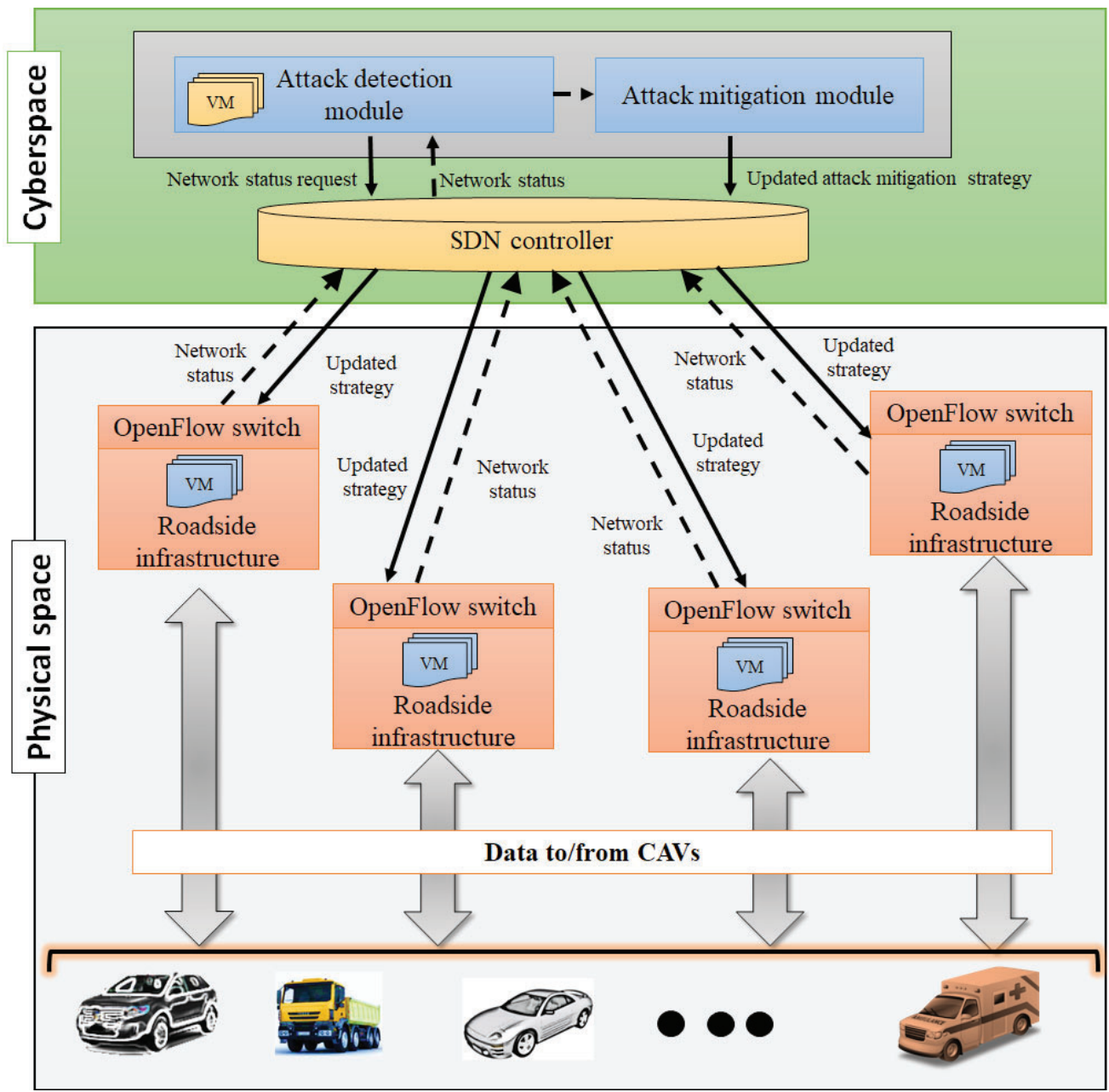

FIGURE 3 Connected and automated vehicle (CAV)–cyberphysical system (CPS) cybersecurity using software-defined network (SDN) and network function virtualization (NFV). VM = virtual machine.

or better performance compared to security modules in proprietary hardware (Han et al. 2015).

In addition, SDN can provide a platform to virtualize network security functions using virtual machines (VMs). With NFV and SDN, security software in a VM can be launched in an SDN-enabled roadside infrastructure within a CAV network (figure 3). SDN and NFV thus act together to improve CAV security in terms of granularity, flexibility, scalability, and resiliency.

*Edge Computing*

Edge computing, in which data are processed in any computing device (e.g., edge device, aggregation centers) close to the data source, ensures high bandwidth usage and distribution of the computational tasks among edge devices (Shi and Dustdar 2016). For example, CAV security modules are deployed near the data source (e.g., in CAVs or roadside devices) for faster processing and effective protection. Edge computing– and AI-based security solutions can be updated/modified in real time to improve cyberattack detection and mitigation. Although the addition of edge devices adds new attack surfaces, edge computing can provide faster security solutions for CAVs by having the solutions closer to CAVs and thus reducing the data transmission delay.

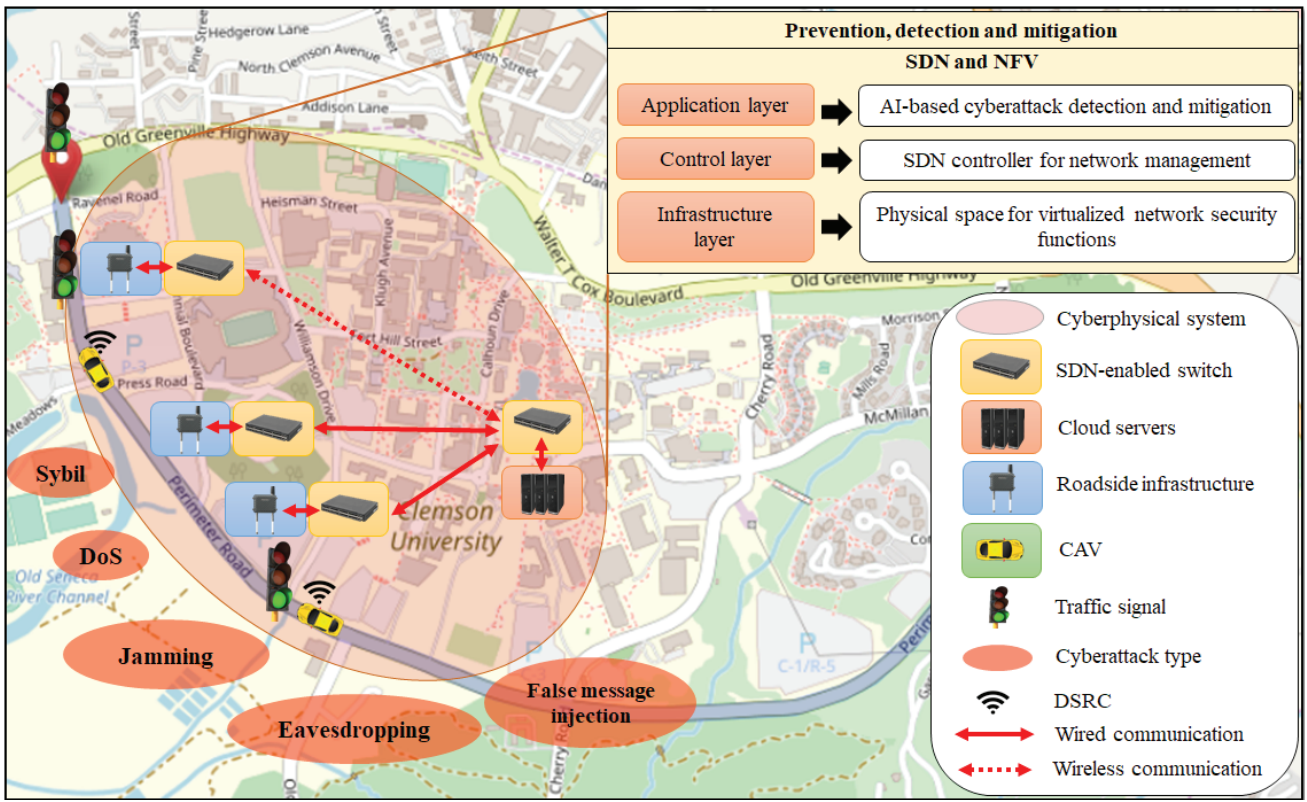

FIGURE 4 Cybersecurity research using edge computing, software-defined network (SDN), and network function virtualization (NFV) at the Center for Connected Multimodal Mobility testbed at Clemson University. CAV = connected and automated vehicle; DoS = denial of service; DSRC = dedicated short-range communications.

At $C^2M^2$ we use an edge computing–based testbed for CAV cybersecurity research (figure 4) (Chowdhury et al. 2018). Different cyberattacks are created against connected vehicles and transportation infrastructure (e.g., traffic signal controllers and roadside devices) to explore effective detection strategies using AI and mitigations using edge computing, SDN, and NFV.

### Information-Centric and Virtual Dispersive Networking

Information-centric networking (ICN) is a new paradigm of communication networking. Unlike traditional host-centric networking, ICN centers on information or content and enables in-network caching and replication; expected benefits include improved network efficiency, scalability, and robustness (Kalogeiton and Braun 2018). ICN in combination with edge computing has the potential to add new dimensions to the network management of connected vehicles (Grewe et al. 2017).

Virtual dispersive networking is another new technology that differs from traditional networks as messages are divided into numerous parts, encrypted, and sent through multiple paths so that attackers can't intercept the whole message, preventing eavesdropping (Twitchell 2013).

### 5G Network

Fifth generation cellular communications are expanding through the Internet of Things (Li et al. 2018). Researchers are adopting 5G in CAVs to pioneer a more secure, reliable, and resilient CAV-CPS (Dey et al. 2018). 5G features such as ultralow latency, higher throughput, and trustworthiness will increase CAV robustness and security compared to the 4G network (Jover and Marojevic 2019).

However, a few 5G protocol specifications are unchanged from 4G, which may transfer some security vulnerabilities to 5G. Moreover, new vulnerabilities are bound to emerge, leading to new security requirements in 5G. The Next Generation Mobile Networks Alliance has identified additional security requirements of 5G wireless networks (NGMN 2015).

Among advantages of the 5G network, edge devices can leverage its higher bandwidth to offload data into the cloud server in real time to facilitate cyberattack detection and mitigation. In addition, 5G features will make it possible to support SDN and NFV in improving network scalability and management. Thus, 5G can make the integration of security solutions easier in a CAV-CPS.

> *Researchers are exploring the use of quantum random number generators to develop ECUs for automotive security.*

### Blockchain

The blockchain is a distributed and trust-based security solution. Using blockchain, CAVs can establish trust with each other, the roadside infrastructure, and cloud servers. The distributed nature of the blockchain makes it difficult to launch an attack as the blocks of data are protected by consensus protocols (Cachin and Vukolić 2017). The blockchain technology may enhance CAV security (Singh and Kim 2017) and privacy (Dorri et al. 2017) by providing a decentralized trusted V2X communication network. However, more research is needed to determine the potential of blockchain technology to improve CAV cybersecurity.

### Quantum and Postquantum Cryptography

Although the development of quantum computers is still in its early stage, quantum computing has the potential to create breakthroughs in AI, optimization, and cryptography (Smedley 2018). One potential use of quantum cryptography for CAVs is quantum key distribution, which can ensure secure key exchange between ECUs in the in-vehicle network (Nguyen et al. 2019). However, this may require an optical fiber connection in the network, which may not be feasible because of the higher production cost (Dennison 2019).

One of the biggest threats to in-vehicle security is the encryption system based on nonrandom numbers. Researchers are exploring the use of quantum random number generators to develop ECUs for automotive security (Nguyen et al. 2019).

As with other technological advances, quantum computing may be used to create CAV cyberattacks. CAVs should leverage postquantum cryptography algorithms to protect against such attacks.

### Conclusions

Researchers in academia and industry are actively working on developing new methods for the detection and mitigation of CAV-CPS cyberattacks. However, challenges persist in this rapidly evolving area.

Given security concerns regarding the acquisition of sensitive information by an attacker, more research should focus on V2X security. CAV data privacy is a concern because CAVs will connect with in-vehicle personal devices and the outside world. Therefore, ensuring V2X security is extremely important to prevent the spread of cyberattacks from CAVs to connected infrastructures and vice versa.

Meeting these challenges requires a concerted effort in academia, industry, and government. Such collaboration can lead to improved standards for a secure CAV-CPS and wide adoption of best security practices across the CAV industry. It will also pave the way for better privacy, safety, and security countermeasures that complement and strengthen each other.

### Acknowledgments

This study is supported by the Center for Connected Multimodal Mobility ($C^2M^2$), a USDOT Tier 1 University Transportation Center, headquartered at Clemson University, South Carolina. Any opinions, findings, conclusions, or recommendations expressed in this article are those of the authors and do not necessarily reflect the views of the $C^2M^2$, and the US government assumes no liability for the contents or use thereof. The authors thank Cameron Fletcher for her time and effort in editing the article thoroughly and improving its quality.